\documentclass[prb,twocolumn,superscriptaddress,floatfix,showpacs,amsmath,amssymb]{revtex4}

\usepackage{graphicx,color}% Include figure files
\usepackage{dcolumn}% Align table columns on decimal point
\newcolumntype{.}{D{.}{.}{-1}}
\usepackage{bm}% bold math
\usepackage[latin1]{inputenc}
\usepackage{epstopdf}
\usepackage{subfigure}

\newcommand{\na}{Na$_3$Ni$_2$SbO$_6$}
\newcommand{\zn}{Li$_3$Zn$_2$SbO$_6$}
\newcommand{\incu}{InCu$_{2/3}$V$_{1/3}$O$_3$}

\newcommand{\bi}{Bi$_{3}$Mn$_4$O$_{12}$(NO$_3$)}

\newcommand{\al}{$\alpha$}

\newcommand{\mbfu}{\(\mu _{\rm B}/{\rm f.u.}\)}

\newcommand{\tn}{$T_{\rm N}$}

\newcommand{\cp}{$c_{\rm p}$}

\newcommand{\cpm}{$c_{\rm p,mag}$}
\newcommand{\cpmm}{$\tilde{c}_{\rm p,mag}$}

\newcommand{\jmk}{J/(mol$\cdot$K)}

\newcommand{\bc}{$B_{\rm C}$}

\newcommand{\bco}{$B_{\rm C1}$}
\newcommand{\bct}{$B_{\rm C2}$}

\begin{document}

%\preprint{APS/123-QED}
\title{Anisotropy governed competition of magnetic phases in the honeycomb quantum magnet \na\ studied by dilatometry and high-frequency ESR}

%%%%  AUTHORS %%%%%%%%%%%%%%%%%%%%%%%%%%%%%%%%%%%%%%%%%%%%%%
\author{J.~Werner}
\affiliation{Kirchhoff Institute of Physics, Heidelberg University, INF 227, D-69120 Heidelberg, Germany}
\author{W.~Hergett}
\affiliation{Kirchhoff Institute of Physics, Heidelberg University, INF 227, D-69120 Heidelberg, Germany}
\author{M.~Gertig}
\affiliation{Kirchhoff Institute of Physics, Heidelberg University, INF 227, D-69120 Heidelberg, Germany}
\author{J.~Park}
\affiliation{Kirchhoff Institute of Physics, Heidelberg University, INF 227, D-69120 Heidelberg, Germany}
\author{C.~Koo}
\affiliation{Kirchhoff Institute of Physics, Heidelberg University, INF 227, D-69120 Heidelberg, Germany}
\author{R.~Klingeler}
\email[Email:]{klingeler@kip.uni-hd.de}
\affiliation{Kirchhoff Institute of Physics, Heidelberg University, INF 227, D-69120 Heidelberg, Germany}
\affiliation{Centre for Advanced Materials, Heidelberg University, INF 225, D-69120 Heidelberg, Germany}

%%%%%%%%%%%%%%%%%%%%%%%%%%%%%%%%%%%%%%%%%%%%%%%%%%

\date{\today}% It is always \today, today,

\begin{abstract}
Thermodynamic properties as well as low-energy magnon excitations of $S=1$ honeycomb-layered \na\ have been investigated by high-resolution dilatometry, static magnetisation, and high-frequency electron spin resonance studies in magnetic fields up to 16~T. At \tn\ = 16.5~K, there is a tricritical point separating two distinct antiferromagnetic phases AF1 and AF2 from the paramagnetic regime. In addition, our data imply short-range antiferromagnetic correlations at least up to $\sim 5\cdot$\tn . Well below \tn , the magnetic field \bco $\approx$ 9.5~T is needed to stabilize AF2 against AF1. The thermal expansion and magnetostriction anomalies at \tn\ and \bco\ imply significant magnetoelastic coupling, both of which associated with a sign change of $\partial L/\partial B$. The transition at \bco\ is associated with softening of the antiferromagnetic resonance modes observed in the electron spin resonance spectra. The anisotropy gap $\Delta = 360$~GHz implies considerable uniaxial anisotropy. We conclude the crucial role of axial anisotropy favoring the AF1 spin structure over the AF2 one. While the magnetostriction data disprove a simple spin-flop scenario at \bco , the nature of a second transition at \bct\ $\approx$ 13~T remains unclear. Both the sign of the magnetostriction and  Gr\"{u}neisen analysis suggest the short-range correlations at high temperatures to be of AF2-type.
\end{abstract}

\maketitle

\section{Introduction}
Mott insulators on layered quasi-two dimensional honeycomb lattices have been found to give rise to a variety of quantum ground states with unusual magnetic excitations. In the $J_1$-$J_2$-$J_3$-model the nature of the ground states maybe of N$\rm \acute{e}$el-, zigzag-, stripe-, or spiral-type and it is governed by the first, second, and third nearest neighbor couplings $J_1$, $J_2$, and $J_3$.~\cite{Fouet2001,Li2012,Cabra2011} In the $S=1/2$ case, the quasi-classical ground states are accompanied by a quantum paramagnetic phase with a valence-bond crystalline type of order where a gapped state is found in the vicinity to the quantum critical points.~\cite{Li2012,Bishop2015} Evaluating the $J_1$-$J_2$-model by a modified spin wave method implies a rich phase diagram, too, including gapped and gapless quantum spin liquid phases.~\cite{Ghorb2016} The recent finding of non-degenerate band-touching points of the magnon bands and Dirac-like behaviour at low energies in theoretical investigations of Heisenberg magnets on a honeycomb lattice in [\onlinecite{Fransson2016}] underlines the need for experimental studies of the low-energy collective spin excitations. However, only few honeycomb systems have been studied by high-frequency electron spin resonance (HF-ESR). In the frustrated $S=1/2$ honeycomb antiferromagnet \incu , HF-ESR has elucidated the peculiar antiferromagnetic ground state.~\cite{Yehia2010,okubo2010high} HF-ESR studies on the $S=3/2$ Heisenberg honeycomb magnet \bi\ suggest that geometric frustration is crucial for suppressing long-range antiferromagnetic order.~\cite{okubo2011anomalous} While in the spin-1/2 case the particular spiral ground state is selected by quantum fluctuations, the spin-3/2 system associated with less pronounced quantum fluctuations is found to yield robust nematic order.~\cite{Mulder2010}

The $S=1$ case realised in \na\ provides further insight into this class of materials as it is a system with small but even spin number. \na\ provides a $S=1$ honeycomb lattice which interlayer coupling is reasonably small, i.e. about 20 times smaller than the leading intraplane exchange.~\cite{Zvereva2015} We observe a tricriticial point in zero magnetic field separating the paramagnetic phase from two antiferromagnetic ones AF1 and AF2 of similar energy. Magnetic fields $B > $\bco\ favor the high-field phase AF2 against the ground state AF1. The $q=0$ low-energy magnon excitations studied by means of antiferromagnetic resonance (AFMR) imply an anisotropy gap $\Delta = 360$~GHz. At $T=4.2$~K, in addition to the field induced phase transition at \bco $\approx$ 9.5~T there is a second one at \bct $\approx$ 13~T. Both transitions affect the dynamic response of the AFMR modes. The magnetostriction data disprove a simple spin-flop scenario at \bco\ but support the picture of anisotropy governed competition of antiferromagnetic phases. Interestingly, both the sign of the magnetostriction $\partial L/\partial B$ and the Gr\"{u}neisen scaling suggest that short-range antiferromagnetic correlations present at least up to $\sim 5\cdot$\tn\ are of AF2-type.
We conclude the crucial role of the axial anisotropy favoring the AF1 spin structure over the AF2 one.

\section{Experimental}

Polycrystalline \na\ was prepared by conventional solid state synthesis as reported in Ref.~[\onlinecite{Zvereva2015}]. Static magnetisation was studied in magnetic fields up to 15~T by means of a home-built vibrating sample magnetometer (VSM) and in fields up to 5~T in a Quantum Design MPMS-XL5 SQUID magnetometer.~\cite{vsm} The relative length changes $dL/L$ were studied on a cuboidal shaped pressed pellet which dimension in the measurement direction is 3.28~mm. The measurements were done by means of a three-terminal high-resolution capacitance dilatometer.~\cite{Wang2009} In order to investigate the effect of magnetic fields, the thermal expansion coefficient $\alpha = 1/L\cdot dL(T)/dT$ was studied in magnetic fields up to 15~T. In addition, the field induced length changes $dL(B)/L$ were measured at various fixed temperatures in magnetic fields up to 15~T and the magnetostriction coefficient $\lambda = 1/L\cdot dL(B)/dB$ was derived. HF-ESR measurements were carried out using a phase-sensitive millimeter-wave vector network analyser (MVNA) from AB Millimetr\'{e} covering the frequency range from 30 to 1000 GHz and in magneict fields up to 17~T.~\cite{Comba2015} For the experiments, a pressed sample pellet of 1.5~mm thickness and diameter of 4~mm was placed in the sample space of the cylindrical waveguide.

\section{Results}

\begin{figure}
\includegraphics[width=1.0\columnwidth,clip] {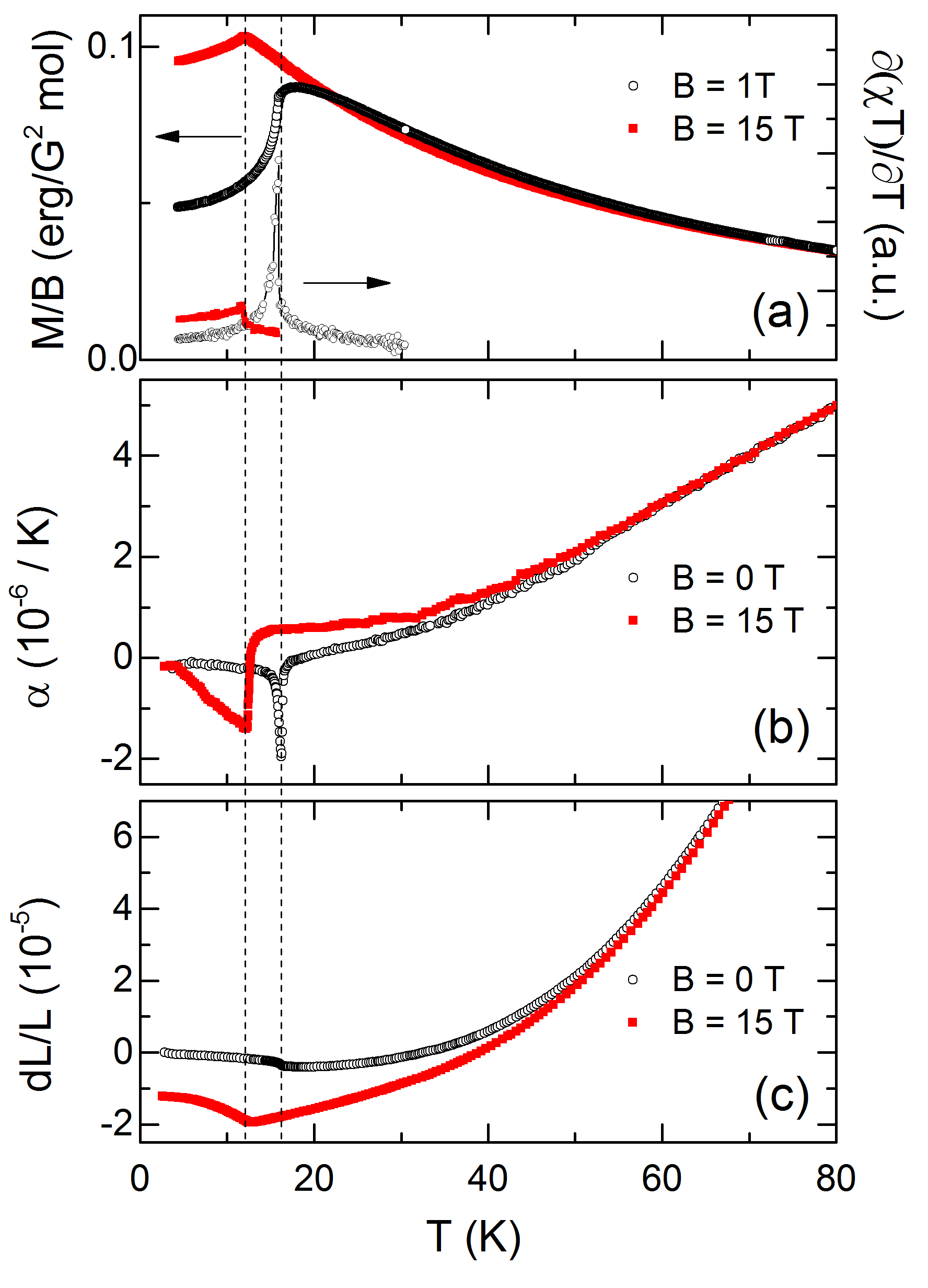}
\caption{(a) Static magnetisation $\chi = M/B$ and magnetic specific heat $\partial (\chi\cdot T)/\partial T$, (b) thermal expansion coefficient, and (c) relative length changes of \na\ vs. temperature, at $B=0$ and 15~T. Vertical dashed lines show \tn\ in the respective field.}\label{alphaL}
\end{figure}

Both the thermal expansion coefficient $\alpha$ (at $B=0$~T) and the magnetic specific heat \cpmm $\propto\partial (\chi\cdot T)/\partial T$ (at $B=1$~T) derived from the static magnetic susceptibility $\chi = M/B$ show a $\lambda$-shaped anomaly signaling the onset of long range antiferromagnetic order at \tn\ = 16.5$\pm$0.5~K (Fig~\ref{alphaL}). The $\lambda$-like anomaly in $\alpha$ is superimposed by a small jump $\Delta \alpha$. Corresponding to the anomalies in $\alpha$ and \cpmm , there is a sharp downturn of the static magnetic susceptibility $\chi = M/B$ and a jump-like increase of the length changes $\Delta L/L \approx 2.2\cdot 10^{-6}$ at \tn . The sign of the anomalies in $\alpha$ and $dL/L$ implies a negative hydrostatic pressure effect on the long range antiferromagnetic order, i.e. $\mathrm{d}T_\mathrm{N}/\mathrm{d}p < 0$.

\begin{figure}
\includegraphics[width=0.95\columnwidth,clip] {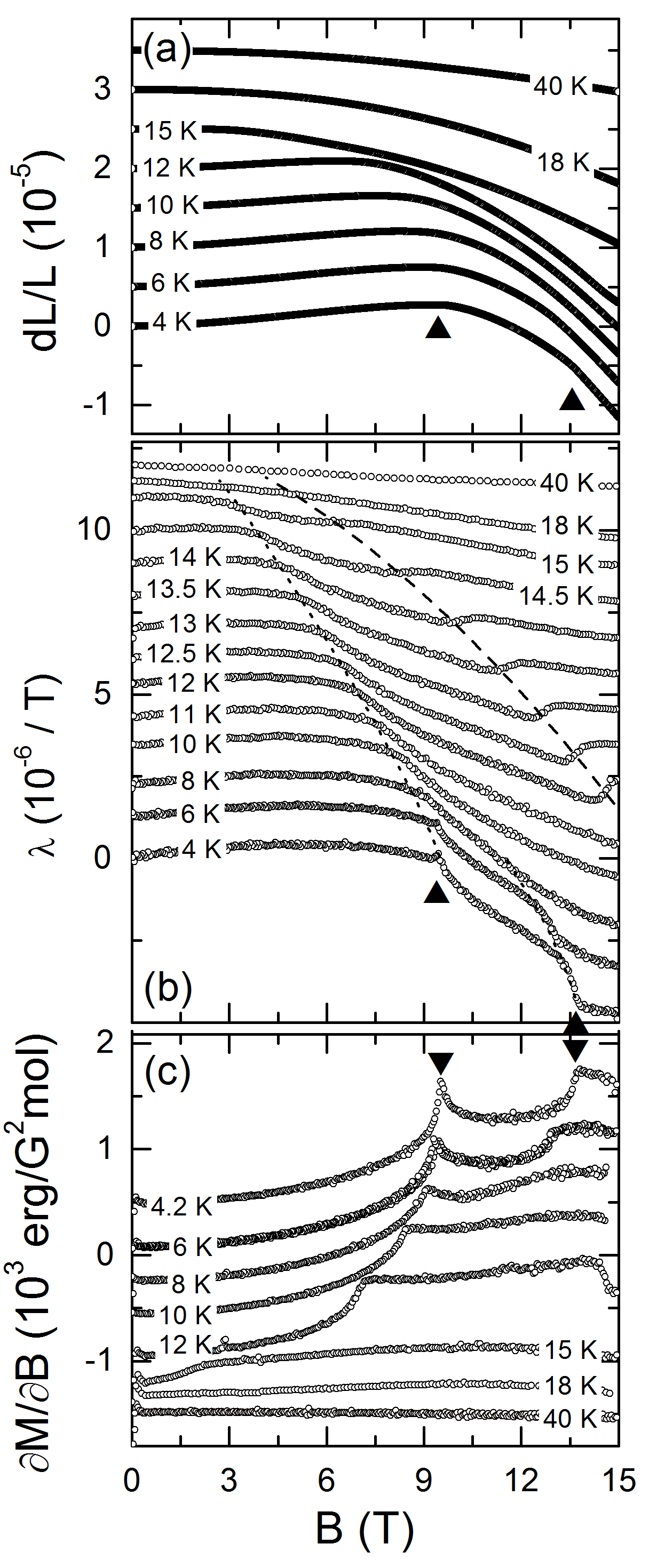}
\caption{(a) Relative length changes, (b) magnetostriction coefficient $\lambda$, and (c) magnetic susceptibility $\chi =\partial M/ \partial B$  of \na\ vs. external magnetic field, at constant temperatures from $T=4$~K to $T=40$~K. The triangles show the two phase transitions at \bco\ and \bct , at $T=4$~K. The dashed line indicates \bc , the dotted lines the $T$-dependence of \bco\ and \bct .}\label{ms}
\end{figure}

Application of external magnetic fields suppresses the long-range antiferromagnetically ordered state as illustrated by the shift of the anomalies in \al\ and \cpmm\ at $B=15$~T in Fig.~\ref{alphaL}. In addition to the shift of \tn , the $\lambda$-like nature of the anomalies changes to a rather jump-like feature. Note, that the length changes $dL/L$ at $B=15$~T shown in Fig.~\ref{alphaL} have been shifted with respect to the zero field data according to the measured magnetostriction at $T=30$~K (see Fig.~\ref{ms}). The data at $B=15$~T imply negative magnetostriction at high magnetic fields in the long range antiferromagnetically ordered phase as well as in a large temperature regime up to about 80~K which is about 5 times \tn . As will be discussed below, at temperatures $T <$ \tn , the behavior at magnetic fields $B\lesssim 10$~T is different. In the derivative of magnetization $\partial M/\partial B$ (see Fig.~\ref{ms}), there is a clear increase associated with application of $B=15$~T at and below \tn , i.e. a left-bending of the magnetisation curve $M$ vs. $B$. In addition, there is a subtle effect above \tn , too, as at $\sim$22~K $\lesssim T \lesssim$ 60~K we observe $\chi$(15~T)$< \chi$(1~T). In this temperature regime, our data hence imply both negative and relatively large magnetostriction and a slight right curvature of the $M$ vs. $B$-curves.

The effect of external magnetic fields is elucidated in more detail in Fig.~\ref{ms} where the relative magnetostriction $dL(B)/L$, the magnetostriction coefficient $\lambda = 1/L\cdot dL(B)/dB$, and the magnetic susceptibility $\partial M(B)/\partial B$ are shown, at various constant temperatures. At $T=4.2$~K, the data show three different phases: At $B<$\bco $\approx$ 9.5~T (phase AF1), there is a positive magnetostriction signaling expansion of the sample. \bco\ is associated with a maximum in $dL/L$, a tiny peak in $\lambda$, and a more pronounced peak in $\partial M/\partial B$. In contrast, \bct\ is associated with a jump-like decrease of $\lambda$ and an associated step in the magnetic susceptibility. Both the intermediate field phase at \bco\ $\leq B \leq$ \bct\ (i.e., AF2) and the high field phase at $B>$\bct\ (i.e., AF3) are characterised by negative magnetostriction with however different values of $\lambda$.

Upon heating up to $\sim$10~K, there is a moderate suppression of \bco\ and \bct . Correspondingly, the peaks in $\partial M/\partial B$ and $\lambda$ transform into kink-like features, the kink at \bco\ being more pronounced than the one at \bct . The low-temperature slopes at 4.2~K of the phase boundaries amount to $dT_{\rm C1}/dB\approx -9.3$~K/T and $dT_{\rm C2}/dB\approx -5.2$~K/T, respectively. At higher temperature, the slope \bco ($T$) becomes steeper and the data indicate a strong suppression of \bco . In addition, at $T\geq 12$~K, there is a step in the magnetostriction coefficient at \bc\ associated with the antiferromagnetic phase boundary \tn ($B$). The well separated phase boundaries \bc\ and \bco\ are particularly evident if the magnetostriction data at 13~K $\leq T \leq 15$~K are considered which exhibit a broad maximum and a step downwards at lower field (signaling \bco ) followed by a step upwards at somehow higher fields (signaling \bc ).

\begin{figure}
\includegraphics[width=0.95\columnwidth,clip] {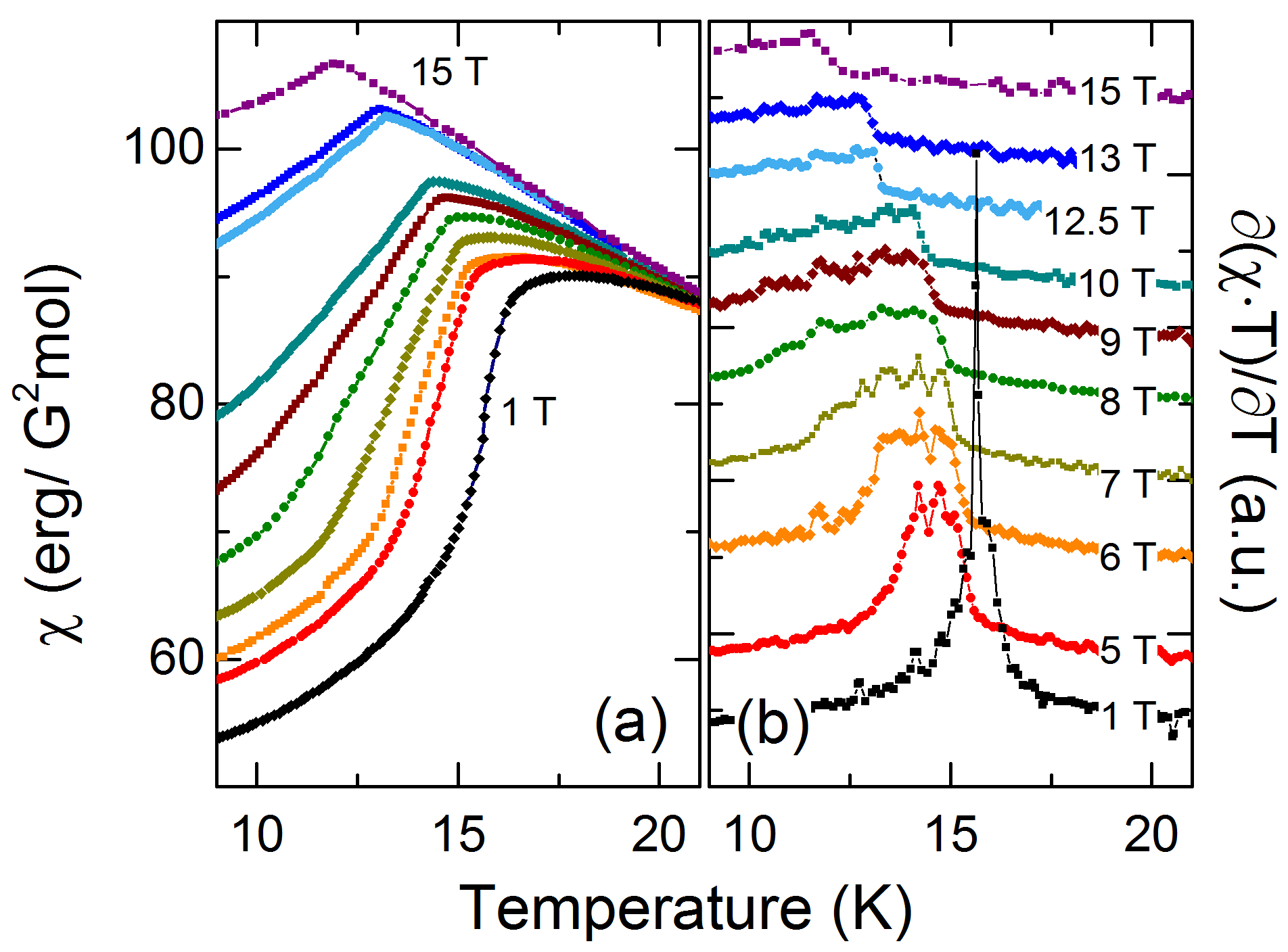}
\caption{Evolution of (a) static magnetic susceptibility and (b) the associated magnetic specific heat $\partial (\chi \cdot T)/ \partial T$  vs. temperature upon application of various constant magnetic fields. Open triangles show $T_{\rm C1}$($B$) (i.e., \bco ($T$)). }\label{mt}
\end{figure}

\begin{figure}
\includegraphics[width=0.99\columnwidth,clip] {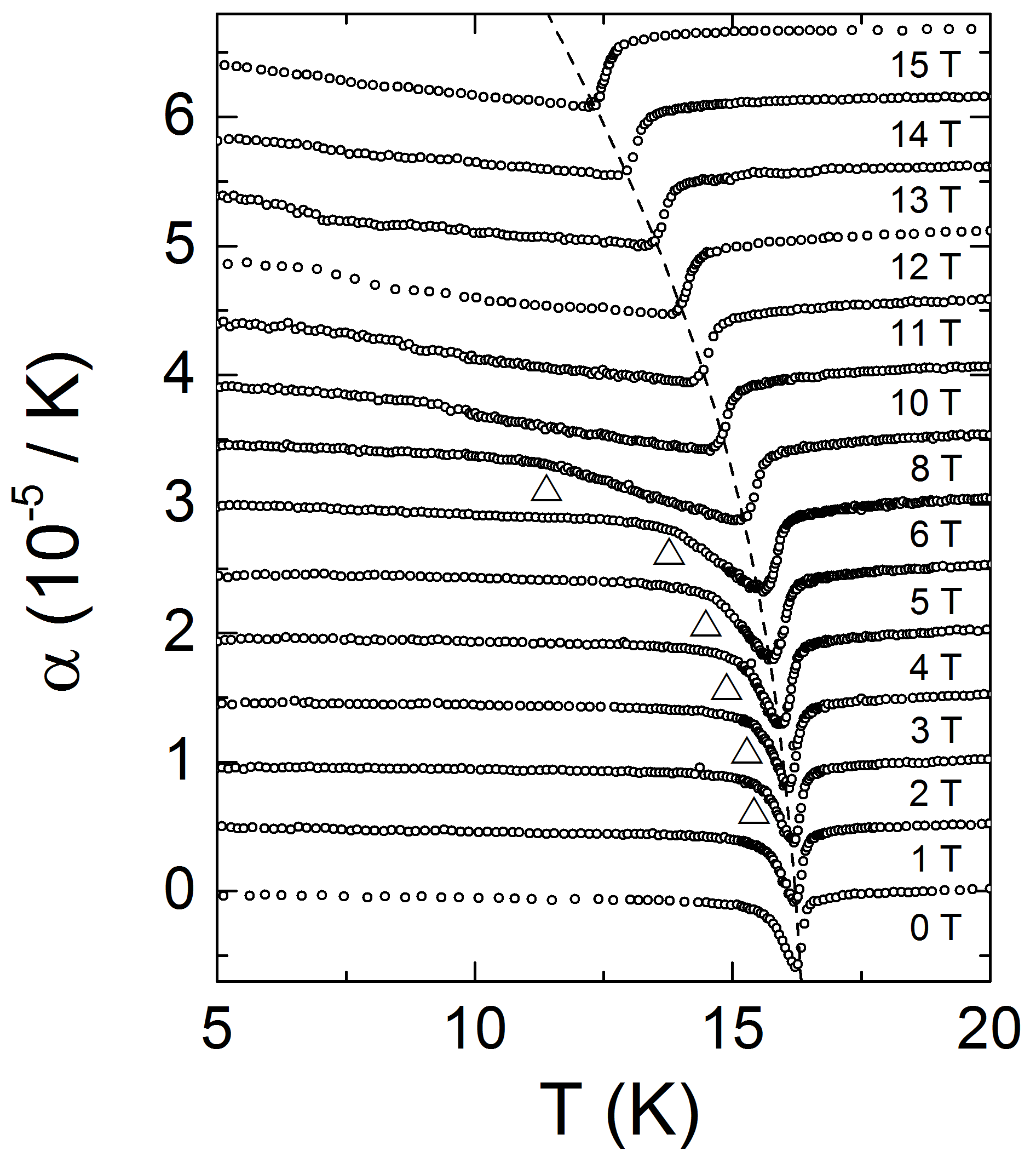}
\caption{Thermal expansion coefficient at various constant magnetic fields (data are shifted). The dashed line indicates \tn ($B$), and the open triangles show $T_{\rm C1}$($B$) (i.e., \bco ($T$)) .}\label{alphafeld}
\end{figure}

The magnetic field effect on both $\chi(T)$ and \cpmm $\propto\partial (\chi\cdot T)/\partial T$, in the vicinity of \tn ($B$), is shown in Fig.~\ref{mt}. Upon application of $B\geq 1$~T, the anomaly in \cpmm\ significantly broadens and covers a regime of, depending on $B$, 2 to 4~K. At high fields, only the high temperature edge of the anomaly is observed as a step. While this step at the right side of the anomalies signals \tn ($B$), comparison with the magnetostriction data suggests that the low-temperature kinks of the broad anomalies at 5~T $\lesssim B\lesssim 8$~T are associated with the phase boundary \bco ($T$). A similar behaviour is observed in the thermal expansion coefficient $\alpha$ (Fig.~\ref{alphafeld}). The anomaly gradually changes from a $\lambda$-type shape to a step-type one (see also Fig.~\ref{alphaL}) while the left shoulder widens. Correspondingly, there is an increasing temperature regime with anomalous length changes which, e.g., at $B=8$~T extends between 15 and 11 K. As will be seen in the magnetic phase diagram, the anomalous length changes signal the AF2 phase appearing between \tn (8~T) = 15~K and $T_{\rm C1}$(8~T) = 11~K.

\begin{figure}
\includegraphics[width=0.95\columnwidth,clip] {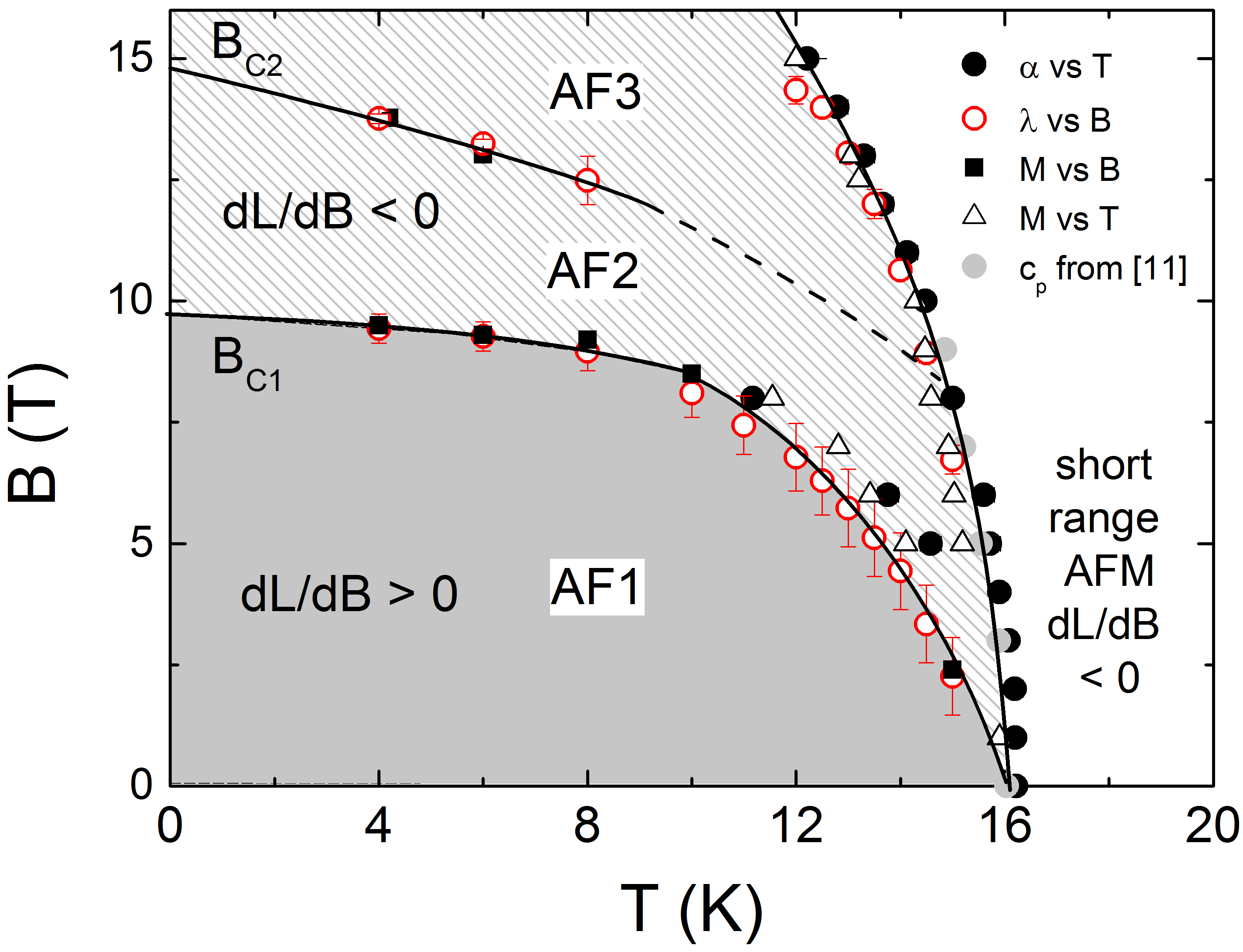}
\caption{Magnetic phase diagram of polycrystalline \na\ as constructed from thermal expansion, magnetostriction, and magnetisation data. AF1 labels long-range antiferromagnetic order with positive magnetostriction $\lambda >0$, AF2 and AF3 indicate antiferromagnetically ordered phases with $\lambda <0$.}\label{PhaseDiagram}
\end{figure}

Summarizing the thermodynamic data yields the phase diagram in Fig.~\ref{PhaseDiagram}. At zero magnetic field, the long range antiferromagnetically ordered phase AF1 below \tn\ is associated with positive magnetostriction. In addition, there is a large regime up to about 80~K where external magnetic fields cause shrinking of the sample volume. As will be corroborated by the HF-ESR data shown below, this regime is characterised by short-range antiferromagnetic order. According to $\partial L/\partial B = -\partial M/\partial p$, positive magnetostriction signals magnetisation decrease upon application of hydrostatic pressure and \textsl{vice versa}. Hence, our data show that, below \tn , application of hydrostatic pressure will be associated with decreasing magnetisation and the opposite behaviour is realised for $T>$ \tn . This observation implies that the short range correlations are not of AF1-type. In addition, the data show a tricritical point at \tn\ which separates  a competing long-range ordered AF2 phase with $\lambda <0$ from the $\lambda >0$ long-range ordered AF1 and the $\lambda <0$ short-range AF ordered phases. At finite magnetic fields $B<$ \bct , this phase appears in-between AF1 and the short-range ordered one. Finally, at $B>$ \bct , there is a third antiferromagnetic phase (AF3), again with $\lambda <0$, which phase boundary is associated by a jump in $\lambda$.

\begin{figure}
\includegraphics[width=1.0\columnwidth,clip] {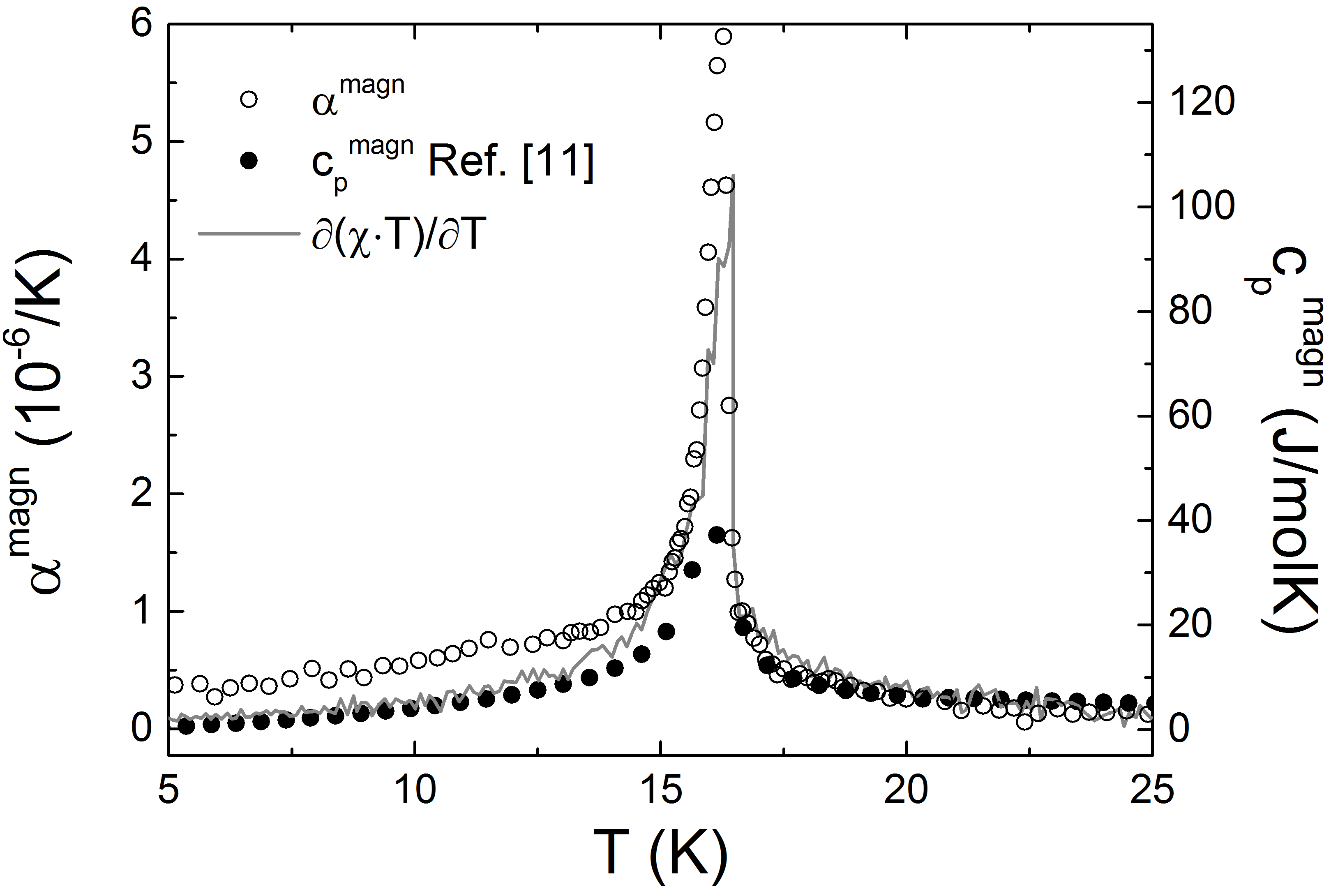}
\caption{Magnetic contribution to the thermal expansion $\alpha^{\rm magn}$ and to the specific heat \cpm\ (from Ref.~[\onlinecite{Zvereva2015}]) as well as the magnetic specific heat \cpmm $\propto\partial (\chi\cdot T)/\partial T$.}\label{gruen}
\end{figure}

Additional information about the nature of the spin ordered phases is obtained by comparing the anomalous, i.e. magnetic, contributions to the thermal expansion coefficient and to the specific heat, $\alpha^{\rm magn}$ and \cpm . Using the Gr\"{u}neisen relation

\begin{equation}
\Gamma = \frac{\alpha^{\rm magn}}{c_{\rm p,magn}}=\frac{1}{V}\left. \frac{\partial \ln \epsilon}{\partial p}\right|_T
\end{equation}

enables determination of the hydrostatic pressure dependence of the associated, i.e. magnetic, energy scales $\epsilon$. The magnetic specific \cpm\ heat has been taken from Ref.~\onlinecite{Zvereva2015}. The phonon contribution to the thermal expansion coefficient is accounted for by scaling the specific heat of the non-magnetic counterpart \zn\ which was used in Ref.~\onlinecite{Zvereva2015} to obtain the magnetic specific heat. As expected, for the phonon contributions to $\alpha$ and \cp\ obtained by this procedure we find a single Gr\"{u}neisen relation in the whole temperature range under study. The resulting volumetric phonon Gr\"{u}neisen parameter amounts to $\Gamma_{\rm phon}=1.9(1)\cdot 10^{-7}$~mol/J. The anomalous contributions to $\alpha$ and \cp , $viz.$ the anomalous magnetic length and entropy changes, are shown in Fig.~\ref{gruen}. For comparison, we also add the magnetic specific heat $\partial (\chi T)/\partial T$ derived from the static susceptibility. The data show that the anomalies in the thermal expansion coefficient and in the specific heat can be scaled to each other at $T>$ \tn\ by applying the  Gr\"{u}neisen parameter $\Gamma_{\rm afm}=-4.6(2)\cdot 10^{-8}$~mol/J. With $V_{u.c.}=108.35~\AA ^3$ [\onlinecite{Schmidt2013}], our analysis yields $\partial \ln \epsilon /\partial p = -3.0(2)\cdot 10^{-3}$/GPa. Assuming $\epsilon$ being proportional to \tn , this corresponds to a small hydrostatic pressure dependence $\mathrm{d}T_\mathrm{N}/\mathrm{d}p = -0.049(3)$~K/GPa. For $T<$ \tn, $\alpha^{\rm magn}$ and \cpm\ do not scale but show the same temperature dependence. To be specific, the same Gr\"{u}neisen parameter $\Gamma_{\rm afm}$ as found for $T\geq T_{\rm N}$ can be used if the step $\Delta \alpha^{\rm magn}$ = $-3.7\cdot 10^{-7}$/K is considered.

\section{High-frequency Electron Spin Resonance}

\begin{figure}
\includegraphics[width=.95\columnwidth,clip] {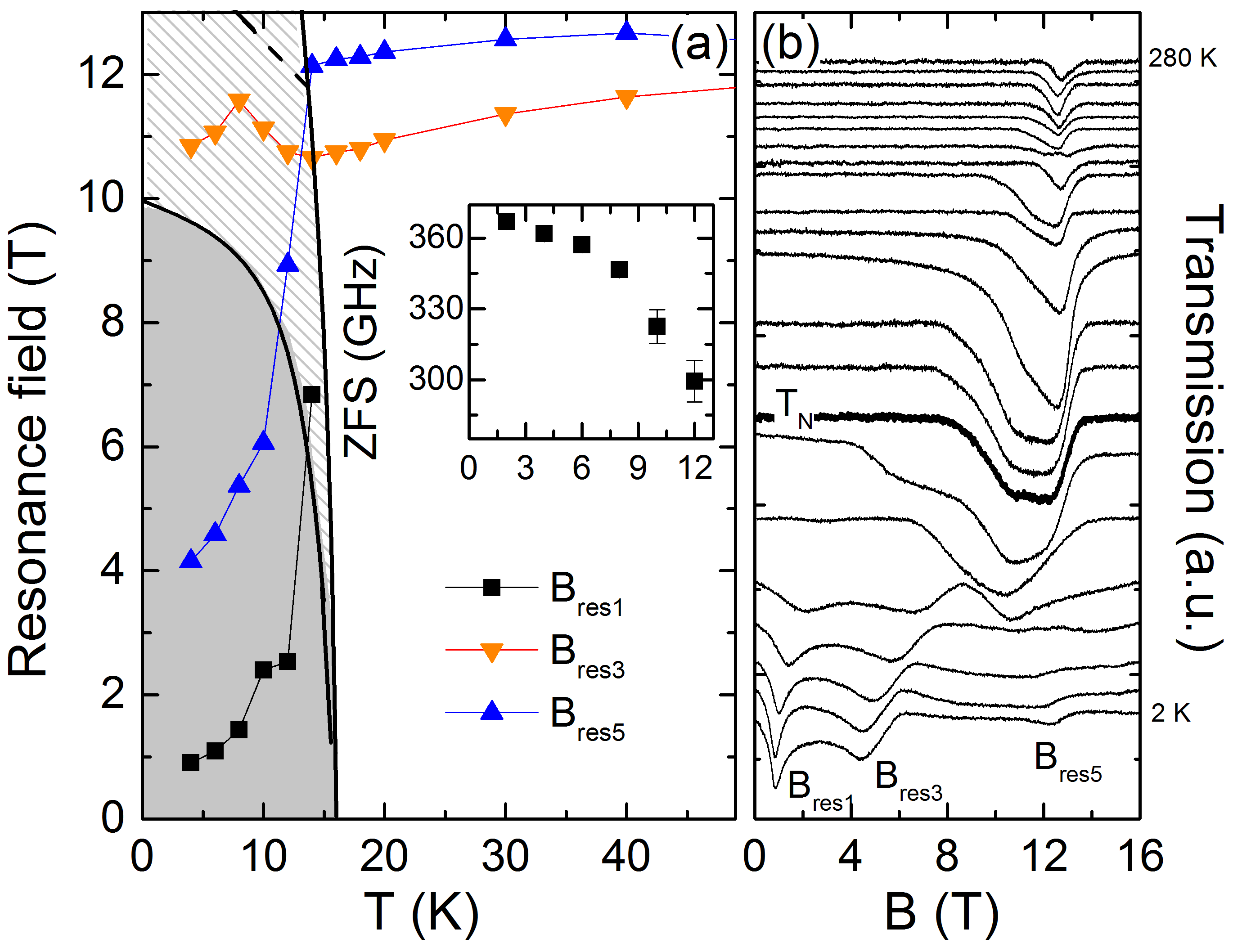}
\caption{Resonance fields at $388$~GHz vs. temperatures between $2$~K and $50$~K (left) and corresponding HF-ESR spectra in the temperature range 2 - 280~K (right). The bold line is the spectrum at \tn ($B=0$). Inset: Zero field splitting (ZFS) vs. temperature.}\label{fig:esr-op}
\end{figure}

ESR spectra taken at $f=388$~GHz (Fig.~\ref{fig:esr-op}b) and at 280~GHz (not shown) show a  broad ($\sim 1$~T) resonance line at $T > 160$~K which position at both frequencies corresponds to the effective $g$-factor of $g = 2.22\pm 0.01$. This value is common for octahedrally coordinated high-spin Ni$^{2+}$ in a paramagnetic phase.~\cite{krzystek2006multi} Upon cooling, the resonance feature shifts and asymmetrically broadens at $T \leq 80$~K. As seen in Fig.~\ref{fig:esr-op}a, below \tn\ in addition to the broadening there is a splitting and considerable shift of the resonance lines which is typically observed in antiferromagnetic resonance (AFMR) spectra due to the evolution of internal fields. In the long range spin-ordered state where HF-ESR is susceptible to collective $q=0$ spin excitations, three AFMR features are observed at $f=388$~GHz and at 280~GHz, respectively (see Fig.~\ref{fig:esr-op}b). With decreasing temperature, the antiferromagnetic resonance lines shift to lower fields for $f=388$~GHz and to higher fields for $f=280$~GHz.

\begin{figure}
\includegraphics[width=1.0\columnwidth,clip] {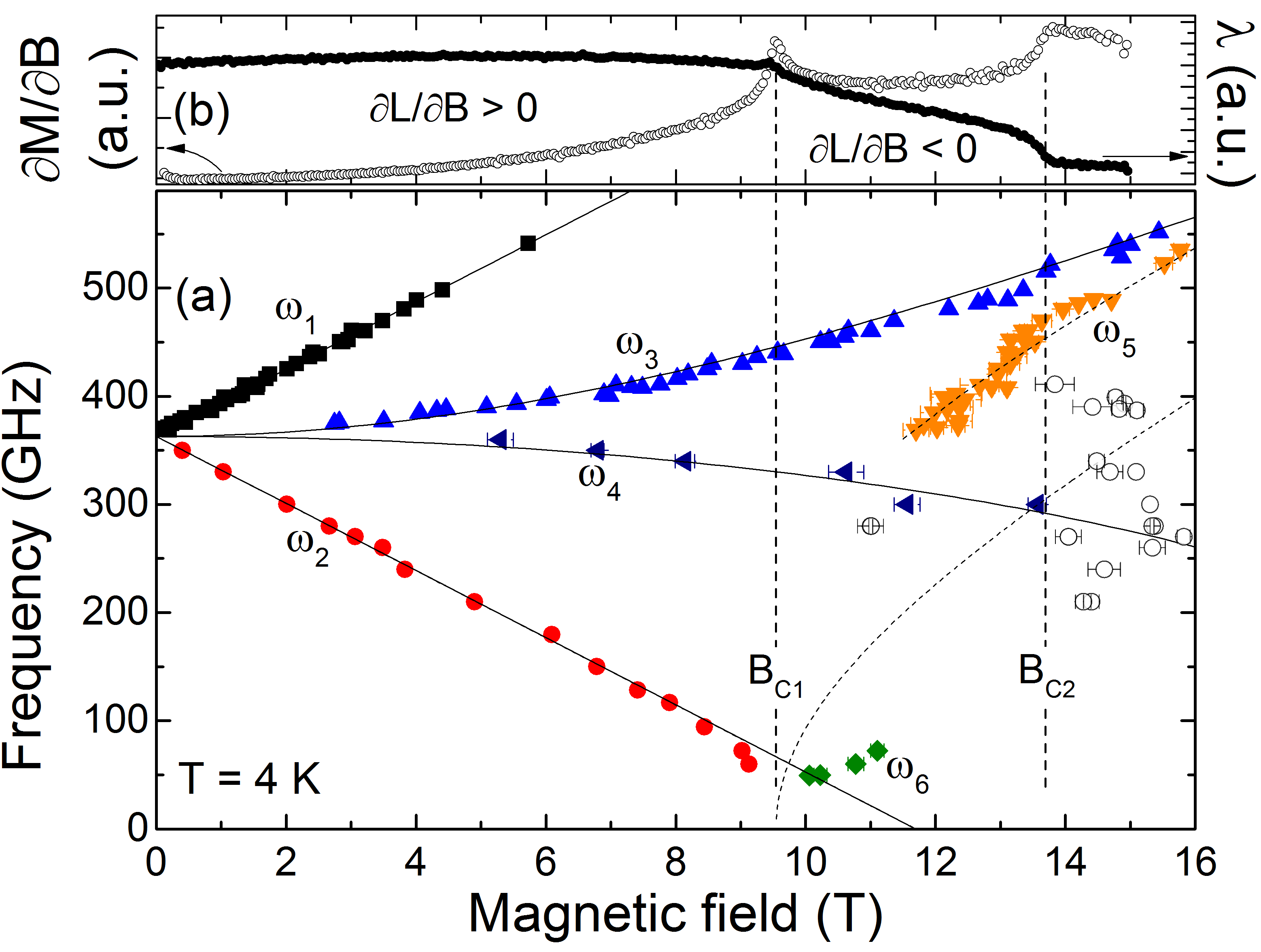}
\caption{(a) HF-ESR absorption frequencies vs. magnetic field at $T=4$~K. Solid lines represent fits according to equations~\ref{eq:AFMR1} to \ref{eq:AFMR3}. (b) Magnetic susceptibility $\partial M/\partial B$ and magnetostriction coefficient $\lambda$, at $T=4$~K, vs. external magnetic field. Vertical dashed lines show the critical fields \bco\ and \bct .}\label{fig:epr-pd}
\end{figure}

HF-ESR measurements at various fixed frequencies enable constructing the magnetic resonance field vs. frequency phase diagram at $T=4$~K as shown in Fig.~\ref{fig:epr-pd}a. The dotted lines correspond to \bco\ and \bct\ from the magnetostriction and the magnetisation measurements (see Fig.~\ref{fig:epr-pd}b). In the AF1 phase, four clearly separated resonances branches $\omega_{1}$ to $\omega_{4}$ are observed. At low fields, all branches merge into a common feature associated with a zero field gap of $\Delta = 360$~GHz. Upon crossing the phase boundary \bco (4~K) = 9.5~T, two of the branches, i.e. $\omega_{3,4}$, remain unaffected while the branches $\omega_{1,2}$ are not observed in AF2. In contrast, a new resonance branch $\omega_5$ is observed in AF2. Neither this new branch $\omega_5$ nor $\omega_3$ show changes at \bct , i.e. when approaching AF3.

The four branches in AF1 suggest a two-sublattice AFMR model with an axial-type anisotropy. Quantitatively, the resonance branches in the uniaxial two-sublattice model are described by the equations \cite{rado1963magnetism}

\begin{align}
\omega_3 & =  \gamma_\perp \sqrt{2B_EB_A+B_A^2+B^2} \label{eq:AFMR1} \\
\omega_4 & =  \gamma_\perp \sqrt{(2B_EB_A+B_A^2) (1-B^2/B_E^2)}
\label{eq:AFMR2}
\end{align}
for the case when the anisotropy field $B_A$ and the external magnetic field $B$ are orthogonal. $B_E$ is the exchange field and $\gamma = g \mu_B$ the gyromagnetic ratio. The case of $B$ parallel to $B_A$ and below $B_{\rm SF}$ is associated with the AFMR modes \cite{rado1963magnetism}

\begin{align}
\omega_{1,2} & =  \gamma_\parallel \sqrt{2B_EB_A+B_A^2}\pm \gamma_\parallel B.  \label{eq:AFMR3}
\end{align}

For the further analysis, we estimate $B_E = 23$~T from the saturation field of the magnetisation.~\cite{Zvereva2015} Fitting of the resonance modes to the data yields an excellent agreement in the AF1 phase (see Fig.~\ref{fig:epr-pd}). We obtain $B_{A}= (1.7\pm 0.5)$~T, $g_\parallel = 2.22\pm 0.01 $, and $g_\perp  =1.98\pm 0.02$. The rather large value of the anisotropy field in comparison to the exchange field underlines the importance of considering the higher order terms in the equations \ref{eq:AFMR1} to \ref{eq:AFMR3}.

In the frame of the uniaxial AFMR model applied here, at $T=4$~K a spin-flop transition is expected at $B_{\rm SF} = (11.7\pm0.5)$~T which is slightly larger than \bco = (9.5$\pm$0.1)~T. We remind that, at \bco , there are structural changes and a sign change of the magnetostriction coefficient which excludes a simple spin-flop nature of the \bco -transition. This conclusion is supported by the fact that the spin-flop mode expected in the two-sublattice model with uniaxial anisotropy at $B>B_{\rm SF}$ is not observed. Neither the parameters obtained in the AF1 phase ($B_{\rm SF} = (11.7\pm0.5)$~T) nor the two-sublattice model employing the (hypothetical) spin-flop field of \bco\ (see the dashed line in Fig.~\ref{fig:epr-pd}a) or any smaller but finite spin-flop field describes the AFMR mode $\omega_5$. As \bco\ is associated with changes of the lattice we can neither exclude a changes of the spin structure at \bco\ nor changes of the anisotropy towards a more complex anisotropy as would be present in a orthorhombic symmetry with several anisotropy axes. Based on the observed mode $\omega_{5}$, however, and without information on the actual spin structure realised in the high-field phases, no clear conclusions on the details on the anisotropy at high fields can be drawn. In the phase AF3 we observe several resonance features in addition to the resonance branches $\omega_3$ and $\omega_5$ in the frequency range 200~GHz to 420~GHz which do not form a resonance branch (open circles in Fig.~\ref{fig:epr-pd}a).

Finally, we analyse the temperature dependence of the resonance fields as seen in Fig.~\ref{fig:esr-op}. The three resonance features showing up in the spectra at 388~GHz refer to the resonance branches $\omega_{1}$, $\omega_{3}$, and $\omega_{5}$. The shift of the resonances upon cooling directly measures the evolution of internal magnetic field, i.e. the internal exchange-anisotropy field strength $B_{EA}=\sqrt{2B_EB_A+B_A^2}$. In the framework of the two-sublattice model, the data hence allow extracting the the antiferromagnetic order parameter $B_{EA} \propto \bigl< S \bigr>$, or the zero field splitting $\Delta = \gamma B_{\rm EA}$. The resulting temperature dependence $\Delta(T)$ of this order parameter is shown in the inset of Fig.~\ref{fig:esr-op}a.

\section{Discussion}

The phase diagram of \na\ shows a tricritical point at \tn . It separates two different long-range antiferromagnetically ordered phases from the paramagnetic one. The latter exhibits short range antiferromagnetic order up to at least 80~K as evidenced by the thermal expansion coefficient and the evolution of the ESR resonance feature. It is further supported by the observation of rather large anomalous magnetostriction in this temperature range.
This direct experimental evidence for short-range correlations extending up to more than $5\cdot$\tn\ agrees to the fact that, below 25 K, only about 60~\% of the total magnetic entropy is released.~\cite{Zvereva2015} The field dependence of the length changes is negative both in AF2 and in the paramagnetic phase while it is positive in AF2. This implies opposite hydrostatic pressure effects, i.e. short-range correlations above \tn\ are not of AF1-type. This conclusion is supported by the Gr\"{u}neisen analysis. Since both AF2 and the paramagnetic, i.e. short-range ordered phase share the same sign of the pressure dependence, we suppose the short-range correlations to be of AF2-type.

The phase boundary \bco ($T$) (or $T_{\rm C1}$($B$)) separating AF1 and AF2 is, at low temperatures, presumingly of weak first order as indicated by a tiny jump in the magnetisation $\Delta M_{\rm C1} \approx 0.12$~\mbfu . Analysing the associated entropy changes by means of the Clausius-Clapeyron relation d$B_{\rm C1}$/d$T$ = -$\Delta S_{\rm C1}$/$\Delta M_{\rm C1}$ yields only small entropy changes of $\Delta S_{\rm C1} \approx 75$~m\jmk\ associated with \bco , i.e. with changing from AF1 $\rightarrow$ AF2 (see the phase diagram in Fig.~\ref{PhaseDiagram}). Indeed we expect only small entropy differences of the spin configurations AF1 and AF2 as both phases develop long range order at the same \tn . This finding somehow agrees to the observation that the Gr\"{u}neisen relation is very similar above and below \tn\ except for a jump in the thermal expansion coefficient $\alpha$. As described above, the short range antiferromagnetic order may be of AF2-type which volume slightly differs from the AF1-type order as shown by the thermal expansion and magnetostriction anomalies. On the other hand, the entropy of the spin configurations AF1 and AF2 is very similar.

Magnetic anisotropy in \na\ is of uniaxial nature which is typical for Ni$^{\rm 2+}$-systems with octahedral coordination of the metal ions. Below \bco , the AFMR modes are well described by means of a two-sublattice model which agrees with the stripe-like spin configuration derived from previous DFT studies.~\cite{Zvereva2015} However, possible competing phases in the $J_1$-$J_2$-$J_3$ honeycomb-lattice model such as the N\'{e}el or zigzag-type phases are not excluded either by our data. In contrast, spiral-type antiferromagnetic phases are not compatible to the two-sublattice description. The high-field phase AF2 however cannot be described in this two-sublattice scenario as the branches $\omega_5$ and $\omega_6$ are not explained. Intriguingly, the branches $\omega_5$ and $\omega_6$ suggest a change in the behaviour at around 11.5~T where no thermodynamic signature of a phase transition is visible. In order to further elucidate these discrepancies, experimental studies of the spin structure in the phases AF2 and AF3 and/or HF-ESR studies on single crystals would be desirable. For example, in the triangular spin-1/2 systems Lu$_2$Cu$_2$O$_5$ the AFMR modes are well described in terms of a six-sublattice model where the modes associated with $B \perp B_A$ are not affected by metamagnetic transitions yielding to a high-field state with cone spin structure.~\cite{goto2000metamagnetic} Without additional knowledge on the spin structure or observation of further branches, however, no reliable model can be constructed for the high field phases.

The phase boundary \bco ($T$) appears at a slightly lower field than expected for the spin-flop transition in the uniaxial two-sublattice model. As the AF1 to AF2 transition is associated with lattice changes there is a weak first order character of the transition. The observed magnetostrictive effect at \bco , the changes in the AFMR modes at \bco , and particularly the changes of the sign of the magnetostriction coefficient $\lambda$ exclude a simple spin-flop nature of the AF2 phase. Instead, AF2 appears to be stabilised when the external magnetic field nearly overcomes uniaxial anisotropy by means of spin reorientation. We hence conclude that spin configuration AF2 is disfavoured with respect to AF1 at small magnetic fields by magnetic anisotropy. Firstly, this is suggested by the sign of the magnetostriction coefficient which excludes the short-range antiferromagnetic fluctuation being of AF1-type but rather suggest a AF2 type. However, while at \tn\ a tricritical point suggests similar energies for AF1 and AF2, well below \tn\ the AF2 phase is stabilised only in magnetic fields which are comparable but slightly below the spin-flop field. Our results hence imply that anisotropy is crucial for stabilisation of the actual antiferromagnetic ground state, i.e. AF1.

\section{Summary}

The phase diagram of $S=1$ honeycomb-layered \na\ exhibits two competing antiferromagnetic phases which display a tricritical point at \tn . The low-energy magnon excitations studied well below \tn\ suggest a magnetic two-sublattice phase AF1 with considerable uniaxial anisotropy showing up in the anisotropy gap of $\Delta = 360$~GHz. Application of external magnetic fields stabilises AF2 at \bco\ which is smaller but of similar order than the anisotropy-exchange field. A simple spin-flop scenario is excluded by the magnetostriction data as well as by the absence of the spin-flop resonance mode. In the paramagnetic phase, both the thermodynamic and the local probe studies show antiferromagnetic fluctuations up to at least $5\cdot T_{\rm N}$. However, short range order is not of AF1-type. We hence conclude a subtle interplay of AF1 and AF2-type spin order which is driven to the actual low-temperature AF1 ground state by magnetic anisotropy.

\begin{acknowledgements}
The authors thank E. Zvereva and A.U.B. Wolter for valuable support, and V.B. Nalbandyan for providing the sample for this study. We thank R. Valenti, V. Kataev, and A. Tsirlin for fruitful discussions. J.W. acknowledges support from the IMPRS-QD.
\end{acknowledgements}

\end{document}